\documentclass[acmsmall, anonymous=false]{acmart}

\usepackage{booktabs} 
\usepackage{graphicx}
\usepackage{epstopdf}
\DeclareGraphicsExtensions{.eps}
\usepackage{amsmath}
\usepackage{bm}
\usepackage{float}
\usepackage{url}
\usepackage{subcaption}

\usepackage[ruled]{algorithm2e} 

\SetAlFnt{\small}
\SetAlCapFnt{\small}
\SetAlCapNameFnt{\small}
\SetAlCapHSkip{0pt}
\IncMargin{-\parindent}

\usepackage{setspace}

\usepackage[colorinlistoftodos,textwidth=2.3cm,textsize=footnotesize]{todonotes}

\newcommand{\sysname}{EdgeXAR\xspace}

\newcommand{\added}[1]{{\leavevmode\color{black}#1}}
\setcopyright{acmcopyright}

\acmDOI{10.475/123_4}
\acmISBN{123-4567-24-567/08/06}
\acmConference[EICS'22]{ACM SIGCHI Symposium on Engineering Interactive Computing Systems }{June 2022}{Sofia Antipolis, France} 
\acmYear{2022}
\copyrightyear{2022}
\acmPrice{15.00}

\begin{document}
\title[\sysname]{\sysname: A 6-DoF Camera Multi-target Interaction Framework for MAR with User-friendly Latency Compensation}
\author{Wenxiao ZHANG}
\orcid{1234-5678-9012-3456}
\affiliation{%
  \institution{The Hong Kong University of Science and Technology}
  \country{Hong Kong}}
\author{Sikun LIN}
\affiliation{%
  \institution{University of California, Santa Barbara}
  \country{USA}
}
\author{Farshid Hassani Bijarbooneh}
\affiliation{%
  \institution{The Hong Kong University of Science and Technology}
  \country{Hong Kong}}
\author{Haofei CHENG}
\affiliation{%
  \institution{University of Minnesota}
  \country{USA}}
  \author{Tristan Braud}
\affiliation{%
  \institution{The Hong Kong University of Science and Technology}
  \country{Hong Kong}}
  \author{Pengyuan Zhou}
\affiliation{%
  \institution{University of Science and Technology in China}
  \country{China}}
  \author{Lik-Hang LEE}
\affiliation{%
  \institution{KAIST}
  \country{South Korea}}
 
\author{Pan HUI}
\affiliation{%
  \institution{The Hong Kong University of Science and Technology}
  \country{Hong Kong}}
\affiliation{%
  \institution{University of Helsinki}
  \country{Finland}}

\begin{abstract}
The computational capabilities of recent mobile devices enable the processing of natural features for Augmented Reality (AR), but the scalability is still limited by the devices' computation power and available resources. In this paper, we propose \sysname, a mobile AR framework that utilizes the advantages of edge computing through task offloading to support flexible camera-based AR interaction. 
We propose a hybrid tracking system for mobile devices that provides lightweight tracking with 6 Degrees of Freedom and hides the offloading latency from users' perception.
A practical, reliable and unreliable communication mechanism is used to achieve fast response and consistency of crucial information.
We also propose a multi-object image retrieval pipeline that executes fast and accurate image recognition tasks on the cloud and edge servers.
Extensive experiments are carried out to evaluate the performance of \sysname by building mobile AR Apps upon it. 
\textcolor{black}{Regarding the Quality of Experience (QoE), the} 
mobile AR Apps powered by \sysname framework run on average at the speed of 30 frames per second with precise tracking of only 1$\sim$2 pixel errors and accurate image recognition of at least 97\% accuracy. As compared to \textit{Vuforia}, one of the leading commercial AR frameworks, \sysname transmits 87\% less data while providing a stable 30\,FPS performance and reducing the offloading latency by 50 to 70\% depending on the transmission medium.
\textcolor{black}{Our work facilitates the large-scale deployment of AR as the next generation of ubiquitous interfaces.}
\end{abstract}

%
%


%
%



\maketitle

\renewcommand{\shortauthors}{Wenxiao ZHANG et. al.}

\section{Introduction}
Augmented Reality (AR) is a natural way of interaction between the real world and digital virtual world~\cite{Yau-subtle}. For a typical AR application, it recognizes the surrounding objects or surfaces and overlays information on top of the camera view with a 3D renderer~\cite{seen-unseen}. Currently, mobile Augmented Reality (MAR) is the most practical AR platform as mobile devices are widely used in our daily life, and many MAR SDKs (e.g., Apple ARKit~\cite{arkit}, Google ARCore~\cite{arcore}, Vuforia~\cite{vuforia}) are released to enable fast development of AR apps.
However, mobile devices are still suffering from the inherent problems of mobile platforms (limited computation power, screen real-estate, and battery life), which restrict their performance in practical AR applications~\cite{ubipoint}.
Most of the AR applications in the market are working in simple scenarios, displaying fixed contents based on detected surfaces, and limiting their usage to gaming or simple technical demonstrations. 

\begin{figure}[t!]
\center 
\begin{subfigure}[t]{0.40\textwidth}
    \includegraphics[width=\textwidth]{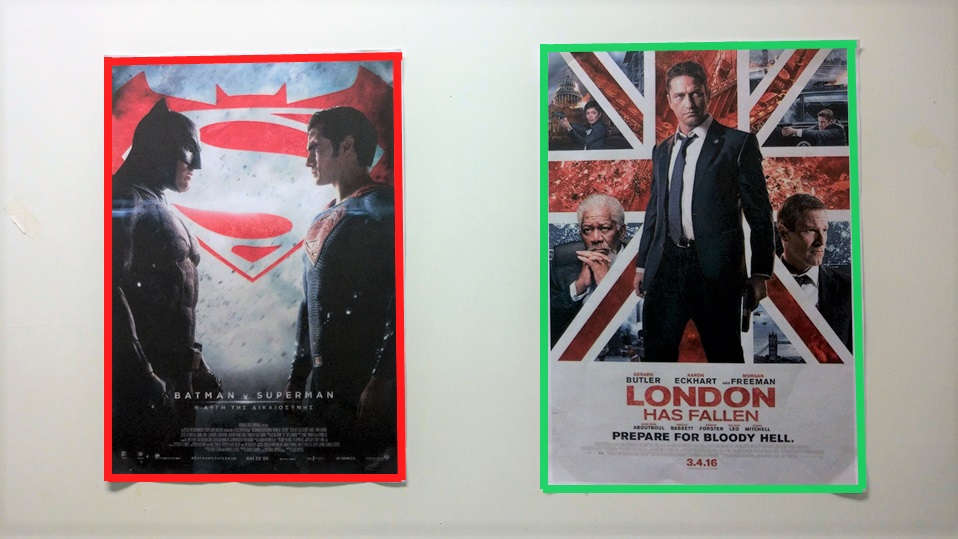}
    \caption{Server recognizes images in the request frame.}
    \label{fig:request}
    \end{subfigure}
\begin{subfigure}[t]{0.40\textwidth}
    \includegraphics[width=\textwidth]{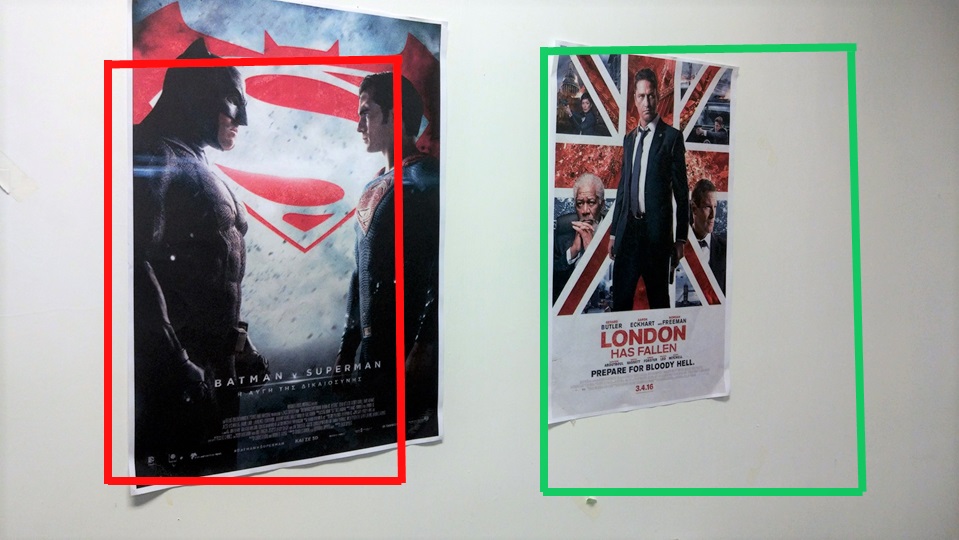}
    \caption{Results mismatch with the current view.}
    \label{fig:result}
    \end{subfigure}
\caption{Alignment problems caused by a high motion-to-photon latency in AR systems.}
\label{fig:mismatch}
\end{figure}

The key enabler of practical AR applications is context awareness, with which AR applications recognize the objects and events in the vicinity of the users and consequently adapts the displayed information to the user's needs~\cite{JamesLamKitYung}.
Large-scale image recognition~\cite{zhu2017discrete,song2017guest} is a crucial component of context-aware AR systems, leveraging the vision of mobile devices with extensive applications in retail, education, tourism, or advertisement. For example, an AR assistant application may recognize road signs, posters, or book covers around users, and overlay useful virtual information on top of those physical elements. 
Despite promising benefits, large-scale image recognition encounters major challenges on mobile platforms. First, large-scale image recognition requires the storage of large image datasets and the corresponding annotation contents are also huge in scale and size. Second, the image recognition task is computation intensive and consumes a lot of time and energy~\cite{yang2015mobile}. \textit{Motion-to-photon} latency is a common evaluation metric in AR systems. It is defined as the latency between an action performed by the user (for instance, a movement), and its actual incidence on the displayed view. With a high motion-to-photon latency, the user is likely to move between the image capture and the completion of the computations. As such, alignment problems arise, where the computation results mismatch the current view of the user. We display example of such alignment problems on Figure~\ref{fig:mismatch}.
It takes on average more than 2 seconds to perform object recognition on a mobile CPU (Snapdragon 800)~\cite{shoaib2015exploiting}. Besides alignment problems, such a high computation latency will result in an extremely low framerate that will further degrade the quality of experience (QoE).


Computation offloading~\cite{flores2018evidence, chatzopoulos2018flopcoin, neto2018uloof, braud2017future} fills the gap between context-aware AR experience and insufficient mobile capabilities.
By utilizing cloud image recognition, cloud-based AR systems 
communicate with the cloud to offload the image recognition tasks and retrieve useful information. 
However, cloud-based systems significantly reduce the computation time at the cost of additional network latency~\cite{ge2017multipath}. 
Edge computing is a promising paradigm that allows to significantly reduce network latency and provides geographical context awareness that simplifies some object recognition requests by reducing the size of the search space. However, contrary to cloud computing, the computing power available is no longer virtually unlimited, and multiple devices compete for a limited -- although more powerful -- computing resource. There is no one-size-fits all solution, and motion-to-photon is thus inevitable in MAR applications, whether on-device, cloud-based, or at the edge of the network.
It is therefore necessary to strike a balance between computation latency and network latency to provide computation results within acceptable deadlines for the user, 
\textcolor{black}{especially when MAR move to the city-wide applications~\cite{lee-csur}}. 
Besides optimising the resource usage, MAR  applications should provide fallback mechanisms to handle outdated results and avoid alignment problems. Existing cloud-based AR systems fail in solving this mismatching problem. They either ask the users to hold their hands steady to coarsely match the result with the current view, or neglect the performance constraints and real-time interaction requirement of mobile devices. 




To address the scalability and latency issues in MAR applications, we propose \sysname, an edge-based MAR framework with both an innovative mobile client design and a powerful image recognition pipeline running on edge servers. \sysname is hardware-agnostic and can run on both CPU and GPU systems.
\sysname solves the alignment problem with an innovative tracking system running on-device, where the historical tracking status of the request frame is recorded and finally utilized by the tracker to compute a corrected position of the physical image after receiving the result. 
By correcting the position of the physical image, our tracking system allows for significant latency compensation, allowing to deploy \sysname not only at the edge, but also on servers located further in the network.
With our fine-grained AR offloading pipeline, AR applications developed with \sysname support large-scale image recognition with the guarantee of high QoE.
To the best of our knowledge, \sysname is the first framework that supports real-time multi-image recognition on mobile clients, overcomes mismatching between outdated offloading result and current view, augments virtual contents according to user's perspective, and enables stable 30 FPS performance for AR applications on off-the-shelf mobile devices. 
Our contribution can be summarized as follows:
\begin{itemize}
\item \textbf{A lightweight mobile tracking system} that provides precise 6 degree-of-freedom (6DoF) multi-object tracking, compensates for mismatching of the recognition result from the user's perspective, and achieves real-time performance at 30 frames-per-second (FPS).
\item \textbf{A multi-object edge image recognition pipeline} that provides low-latency, high-accuracy multi-image recognition, as well as pose estimation services.
\item \textbf{A fine-grained edge offloading pipeline} that minimizes the overall recognition latency and mobile energy consumption in mobile scenarios.
\item \textbf{Extensive evaluation} of \sysname for both edge and cloud computing scenarios. \sysname allows for real-time 30\,FPS multiple image recognition and tracking with only 1\textasciitilde 2 pixel error. It performs tracking under 25\,ms for up to 240 feature points and can compensate offloading latency up to 600\,ms. Compared to leading commercial cloud AR recognition frameworks such as \textit{Vuforia}, \sysname provides a constant 30\,FPS framerate, transmits 87\% less data over the network, and reduces the oflloading latency by 50\% in WiFi and 70\% using mobile broadband networks.
\end{itemize}


The remainder of this paper is organized as following:  Section~\ref{sec:relatedwork} summarizes the existing work on MAR systems and latency compensation techniques. Section~\ref{sec:designchoice} discusses the design choices of \sysname. Section~\ref{sec:systemDesign} gives an overview of the system architecture. Section~\ref{sec:clientDesign} and~\ref{sec:serverDesign} describe the mobile client design and server design, respectively. The implementation details and the experimental results are shown in Section~\ref{sec:evaluation}. Finally, the conclusion is given in Section~\ref{sec:conclusion}.

\section{Related Work}
\label{sec:relatedwork}

As a complete framework for MAR applications, \sysname combines elements of on-device MAR, cloud and edge-based MAR, image retrieval techniques, and latency compensation strategies. In this section, we summarize the main works related to these elements.

\subsection{On-device Mobile Augmented Reality}

Typical mobile AR SDKs can be divided into two categories: a) traditional SDKs (e.g., Wikitude~\cite{wikitude}, Vuforia~\cite{vuforia}, Blippar~\cite{blippar}) that heavily utilize computer vision technologies for marker-based AR. These SDKs first recognize and track the relative camera pose to the marker for each frame; b) emerging AR SDKs (e.g., Google ARCore~\cite{arcore}, Apple ARKit~\cite{arkit}) that bring marker-less AR onto the mobile platform. Such markerless SDKs calculate an absolute pose of the camera in the environment with visual-inertial odometry (VIO)~\cite{li2013high,leutenegger2015keyframe}.

For marker-based AR, Nate et al.~\cite{hagbi2009shape} proposes a mobile system to recognize and estimate the 6DoF pose of planar shapes at interactive frame rate using recursive tracking.
Wagner et al. applies SIFT and Ferns to mobile platforms with proper modifications~\cite{wagner2010real}.
They also built a mobile system for multi-object tracking and detection~\cite{wagner2009multiple}. Global object pose estimation and device localization are also implemented on the mobile platform~\cite{ventura2014global,arth2015global}. Closer to our solution, Amateur~\cite{10.1145/3287033} performs matches navigation information and road-view vision on-device and provides compensation mechanisms for incomplete and erroneous detection.

For marker-less AR, both ARCore and ARKit leverage feature points from the monocular camera and motion data from the inertial measurement unit (IMU) to track the pose of the mobile device in 3D. Their tracking capability is more flexible and scalable compared to marker-based AR. 

Although these solutions allow the development of MAR applications without requiring external computation capabilities, they are limited by the capabilities of the device. As such, these systems can only operate within a constrained context, and thus present limited performance and scalability. On the other hand, \sysname combines edge computing with on-device computations for real-time and scalable MAR applications.

\subsection{Cloud- and Edge-based Mobile Augmented Reality}
Google Goggles~\cite{goggles} looks up the database to recognize a picture taken by user and retrieve useful information.
The systems in~\cite{pilet2010virtually,gammeter2010server} address the scalability issue by integrating image retrieval techniques into the tracking system on a PC.
However, their work cannot handle the mismatching problem.
\cite{jung2012efficient} describes a similar system, in which extra information like region of interest (ROI) of the object is required to initialize the recognition and tracking. 
Overlay~\cite{jain2015overlay} requires a user to hold the camera still to display annotation in a coarsely correct position.
VisualPrint~\cite{jain2016low} is another cloud-based mobile augmented reality system which uploads extracted features instead of raw images or video stream to save network usage. Finally, cloud-based AR systems also allow to introduce further features, such as collaborative creation, synchronously or asynchronously~\cite{10.1145/3351241}. 
Compared to above cloud-based AR systems, \sysname handles the offloading latency by design, and compensates for mismatching issues through on-device object tracking. \sysname also minimizes the overall latency and energy consumption.

Other systems offload only the most computation-heavy operations to the cloud.
Jaguar~\cite{zhang2018jaguar} is an application that leverages GPU systems at the edge of the network to perform object recognition. In comparison, \sysname is a hardware-agnostic framework that can provide real-time performance with servers located in the cloud or at the edge thanks to its latency compensation capabilities. The tracking system in \sysname allows for a finer accuracy, with only 1-2 pixel error.  CloudAR~\cite{zhang2017cloudar} proposes dynamic offloading of computations in the cloud. \sysname is able to leverage both edge and cloud servers, and proposes advanced latency compensation techniques to account for the additional offloading latency.

\subsection{Image Retrieval Techniques}
One of the fundamental topics in image retrieval is the image feature detection and extraction.
SIFT~\cite{lowe1999object} and SURF~\cite{bay2006surf} provide good results with robust scale and rotation invariant.
CNN features~\cite{krizhevsky2012imagenet} provide state-of-the-art representation accuracy.
However, these approaches are relative slow. Combined with binary descriptors, corner feature points detectors such as Harris corner detector~\cite{harris1988combined}, FAST~\cite{rosten2010faster}, and AGAST~\cite{mair2010adaptive} provide faster solutions.

One can encode images into single vectors for further image retrieval and classification. 
Authors in~\cite{csurka2004visual} introduce the bag-of-visual-words (BOV) model.
The authors in~\cite{jaakkola1999exploiting,perronnin2010improving,jegou2012aggregating} present Fisher kernel encoding to improve the representational accuracy in comparison with k-means clustering used in BOV.
In recent years, Fisher Vector (FV) has been widely used in image retrieval with good performance as shown in~\cite{perronnin2010large,douze2011combining}. 
The authors in~\cite{indyk1998approximate,gionis1999similarity} propose LSH for fast nearest neighbor search, and many improvements are proposed in~\cite{datar2004locality,andoni2006near,kulis2009kernelized}. 
Some other image feature coding and hashing techniques are proposed in~\cite{shen2017asymmetric,ercoli2017compact}.

\sysname combines multi-scale AGAST detector with FREAK binary feature descriptors. We finally do Fisher encoding on binary features for image recognition. 

\subsection{Latency Compensation Techniques}
Volker Strumpen et al. presentes a latency hiding protocol for asynchronous message passing in UNIX environments~\cite{Strumpen1995}. With this protocol distributed parallel computing can be utilized in applications.

Outatime~\cite{lee2015outatime} is a mobile cloud gaming system which delivers real-time gaming interactivity. The basic approach combines input prediction with speculative execution to render multiple possible frame outputs which could occur in a round trip time of future. 
Kahawai~\cite{cuervo2015kahawai} is another mobile gaming systems which aims at providing high-quality gaming quality by offloading a portion of the GPU computation to server-side infrastructure. In contrast with previous thin-client approaches which require a server-side GPU to render the entire content, Kahawai uses collaborative rendering to combine the output of a mobile GPU and a server-side GPU into the displayed output.


Different from those existing methods, \sysname specifically targets latency compensation for MAR, which is more challenging as 
the alignment problem between the virtual and the physical world is more sensitive to latency, while being less predictable. 
 \sysname provides seamless real-time user experience while keeping the mobile client as lightweight as possible.

\section {Background and Design Choices}
\label{sec:designchoice}

\begin{figure*}[t]
    \centering
    \includegraphics[width=1\textwidth]{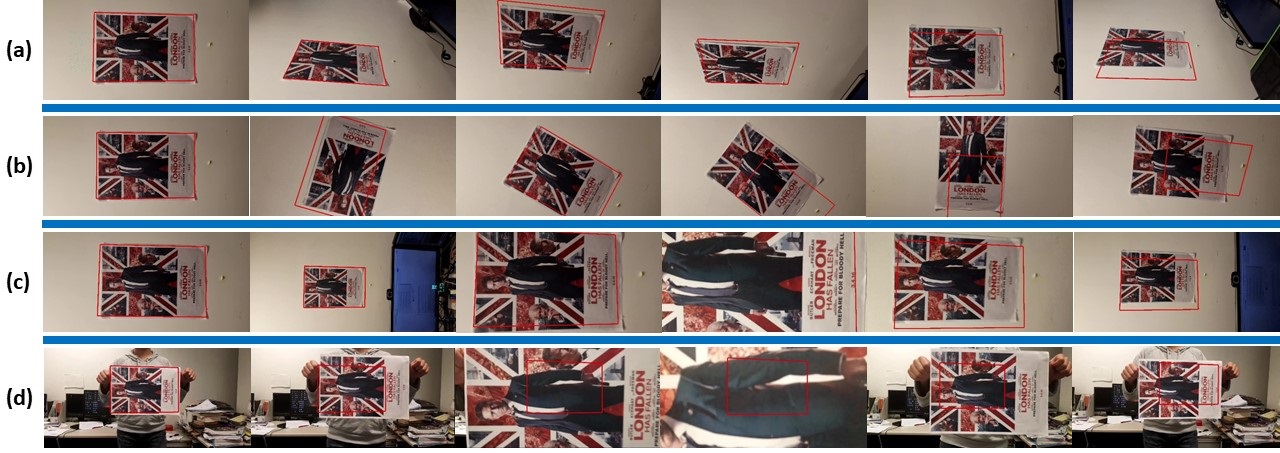}
    \caption{Results of ARCore tracking in four scenarios: (a) view changing; (b) camera rotation; (c) scaling caused by forward and backward movement; (d)object moving. The red edges are the tracking results on the mobile client.}
    \label{fig:arcore}
\end{figure*}

In this section, we explore the design space of the cloud and edge-based AR systems, and discuss the advantages and disadvantages of possible solutions to motivate the design of \sysname.

\subsection{Is VIO a Good Solution for Context-aware AR?}

ARKit and ARCore hit the market, and the technology underneath, VIO locates mobile devices in the environment independent of the detection of specific markers, providing a more flexible and scalable tracking capability when compared with traditional marker-based AR. However, VIO cannot recognize the surrounding objects and get context information. Most of existing AR apps powered by ARCore or ARKit  work on the detected horizontal surfaces (e.g., table, floor) for gaming or demonstration. We have to make the decision between marker-based tracking and marker-less tracking.

To learn the tracking performance of ARCore, especially in the scenario of tracking recognized image, we build an AR app based on ARCore, which recognizes one image stored in the app and tracks it afterwards. The app is deployed on a Samsung Galaxy S8 smartphone, which is a flagship model and well-calibrated for ARCore. The tracking results of four representative sequences are shown in Figure \ref{fig:arcore}, from which we find several problems of VIO-based AR systems:

\textbf{VIO tracking of markers are inaccurate and unstable.} As shown in Figure \ref{fig:arcore}, ARCore tracking results are not accurate in the cases of camera tilt, rotation and scaling, with obvious mismatching between the physical image and the tracking result. Unstable tracking results arise frequently as well, where the annotated image edges hop during tracking as shown in the second half of sequence (b), and this is not a rare case as hopping happens in four out of ten repeated attempts. As VIO tracking combines both the visual and inertial outputs, the noisy and biased inertial measurements are the main reason for this unsatisfying performance.

\textbf{VIO needs good initialization.} Upon loading of the app, a initialization procedure is needed before the tracker would work. The initialization usually takes few seconds, and the user has to hold and move the phone for a few seconds. On the contrary, visual-based tracker used in marker-based AR would start working instantly upon loading without an initialization.

\textbf{VIO cannot track moving objects.} Unlike the tracking of object's relative pose in visual tracking systems, typical VIO systems are supposed to work in a rigid environment, where the localization procedure locates the devices' pose and ignores moving objects, as shown in sequence (d) of Figure \ref{fig:arcore}. Although some existing work have addressed the moving object tracking problem~\cite{wang2003online,wang2007simultaneous} in simultaneous localization and mapping (SLAM)  systems, these methods make major modifications of the SLAM pipeline, which add up to the overall complexity and is not currently supported by ARCore or ARKit.

\textbf{VIO needs physical size of recognized image.} Without considering the dynamic object tracking, the static object tracking of VIO is different from that of visual tracker as well. Visual tracker works with the relative pose and size of the physical object and annotation contents, so that the remote recognition service would return the pose of the image object without caring about the real size of the image. However, as the odometry unit of VIO systems is in actual physical size, the tracking of static objects needs the physical size of the image as well, which is not possible to infer on a remote server and not trivial to get on the mobile client. In this app developed with ARCore, we have to manually input the physical size of the image, which limits its pervasive usage. 

With the shortages shown above, a VIO based solution is not adequate in context-aware AR with image recognition functionalities. Marker-based AR performs better in terms of accuracy, robustness, ease of use, and availability for image tracking at the current stage.

\subsection{Marker-based AR Pipeline}

\begin{figure}[t]
    \centering
    \includegraphics[width=\linewidth]{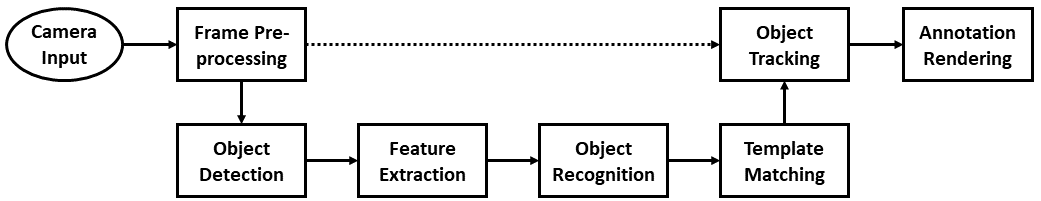}
    \caption{Typical mobile AR pipeline.}
    \label{fig:arpipeline}
\end{figure}

A typical pipeline of marker-based mobile AR systems has 7 building blocks, as shown in Figure \ref{fig:arpipeline}. It starts with Frame Preprocessing that shrinks the size of a camera frame, e.g., by downscaling it from a higher resolution to a lower one. The next step is Object Detection that checks the existence of targets in the camera view of a mobile device and identifies the regions of interest (ROI) for the targets. It will then apply Feature Extraction to extract feature points from each ROI and Object Recognition to determine the original image stored in a database of to-be-recognized objects. Template Matching verifies the object-recognition result by comparing the target object with the recognized original image. It also calculates the pose of a target (i.e., position and orientation). Object Tracking takes the above target pose as its initial input and tracks target object between frames in order to avoid object recognition for every frame. Finally, Annotation Rendering augments the recognized object by rendering its associated content.

\subsection{Latency Compensation Choices for User-centric AR Systems}

\begin{table}
  \begin{tabular}{lll}
    \toprule
    Template Matching & Mobile Client & Edge Server  \\
    \midrule
    Time Consumption(ms) & $438.2\pm96.5$ & $57.62\pm6.14$\\
    Energy Consumption(uAh) & $276.0\pm60.9$ & N/A \\
    \bottomrule
  \end{tabular}
\caption{Time and energy consumption for template matching with SIFT features on a Xiaomi Mi5 phone and an edge server, respectively. The value after $\pm$ is the standard deviation (for 20 experimental runs).}
\label{tab:templatematching}
\end{table}

As offloading image recognition to a remote location, whether at the edge or in the cloud, will bring inevitable latency, edge and cloud-based AR systems need to handle this latency properly so that the annotation contents can be aligned with the physical objects precisely. For marker-based AR, there are two ways to hide the latency.

One possible solution is that the server sends the original image back and the client executes template matching to calculate the current pose of the image. However, template matching is not a trivial task on mobile, which consumes much time and energy when compared with edge execution, as shown in Table \ref{tab:templatematching}. Executing template matching on the server reduces the overall pipeline latency significantly. On the other hand, even if template matching is executed on the client, the time consumption of template matching would also make the mismatching happen, as several frames have passed during the template matching procedure.

Another solution is tracking the image movement during the edge recognition procedure, as tracking is much more light-weight compared to template matching. With the tracked image pose transformation during the edge recognition procedure, the recognition result can be corrected accordingly to match the current view. A previous work Glimpse~\cite{chen2015glimpse} proposes to solve the mismatching problem with tracking as well, but Glimpse caches selected frames during edge recognition and starts tracking after receiving the result, with the recognition result utilized as the tracking initial input. However, this design would erupt processing burden on the mobile device, and the tracking of a sequence of frames takes long time and causes mismatching as well. 

As a result, our system disperse the tracking processing during the edge recognition procedure. Our design achieves overall low latency while keeping the mobile client lightweight and fluent.

\section{System Design Overview}
\label{sec:systemDesign}

\begin{figure}[t]
    \centering
    \includegraphics[width=\linewidth]{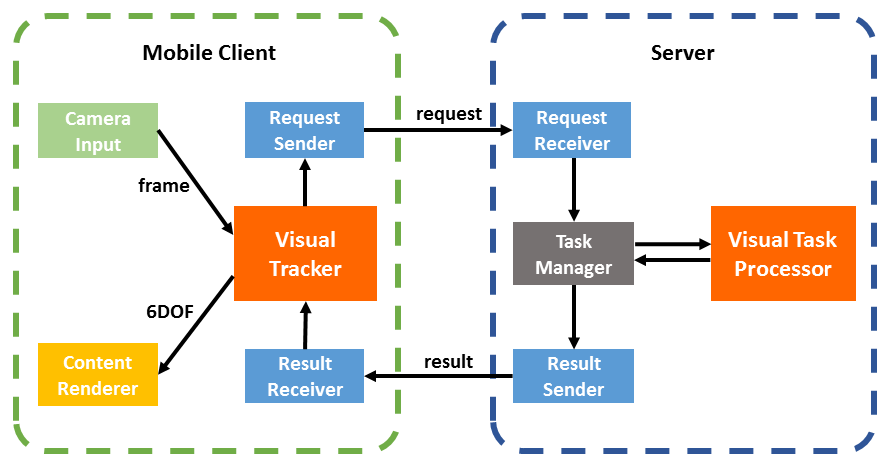}
    \caption{\sysname overview showing the data flow and the main components of the mobile client and the server.}
    \label{fig:overview}
\end{figure}
Figure \ref{fig:overview} shows an overview of the system architecture, that includes the mobile client and the edge server. 
Following is a brief description of the overall procedure: 

On the mobile client, visual tracker gets the camera video feed and starts extracting and tracking feature points within view once the application is launched. Unlike~\cite{jung2012efficient}, our system would not require the user to input the region of interest (ROI), and camera stream is the only input of the whole framework. Therefore, the visual tracker simply tracks feature points of the entire frame at this time. An object recognition request is sent to server to recognize target objects inside the camera frame. 

The server creates worker threads to process the received request. These worker threads handle the receiving and sending messages on the network, as well as performing the visual tasks. 

The result is then sent back to the mobile client, including poses and identity information of the recognized objects. With feature points tracking, visual tracker is able to calculate the location transition of any specific area inside the frames. Therefore the current poses of the objects can be derived from the result, which are poses of objects within the request frame. These poses are further utilized by the content renderer to render virtual contents. 

This process repeats periodically under visual tracker's scheduling, thus new objects are recognized and augmented continuously.

\section{Mobile Client Design}
\label{sec:clientDesign}
The mobile client works mainly on tracking and contents rendering while offloading the heavy image recognition tasks to an edge server. 
In this section, we focus on introducing the functional modules on mobile client. 

\subsection{Visual Tracker}

\begin{figure*}[t]
    \centering
    \includegraphics[width=0.7\textwidth]{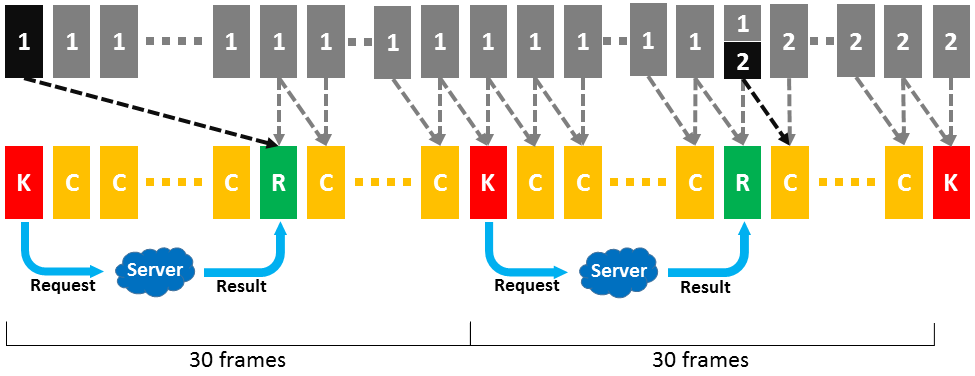}
    \caption{Visual tracker pipeline. The sequence above in grayscale stands for the feature point extraction and tracking pipeline, where each number indicates the feature points set number; the sequence below in color stands for the objects pose initialization and update pipeline, where K stands for a key frame, C stands for a continuous frame, and R stands for a result frame.}
    \label{fig:frames}
\end{figure*}

\begin{figure*}[t]
    \centering
    \includegraphics[width=1\textwidth]{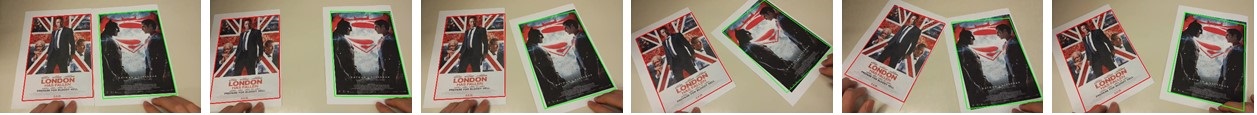}
    \caption{Multi-object tracking of \sysname. The red and green boundaries are the output of the visual tracker, allowing users to perform camera-based AR interaction with objects with high flexibility.}
    \label{fig:multitrack}
\end{figure*}

Visual tracker (Fig.~\ref{fig:frames}) carries out the core logic of mobile client as a   scheduler of other modules. We design our own feature point based tracker for two reasons. First, \sysname starts tracking without a ROI or the object position, and the renderer needs precise 6DoF results rather than simple rectangle bounding boxes to position virtual contents in a 3D space. Second, visual tracker must achieve a high FPS on the mobile client, and feature point based tracking (e.g., optical flow~\cite{lucas1981iterative}) is lightweight enough to fulfill this requirement.

We decouple the feature tracking and object pose estimation tasks into two separate pipelines, as shown in Figure \ref{fig:frames}. The first pipeline works on feature point extraction and tracking for the whole frame. The second pipeline works on pose initialization and recognized objects update.

\textbf{The feature point extraction and tracking pipeline,} which is the basis of visual tracker, starts processing the camera frames upon launching the mobile client. 
At the beginning, the pipeline extracts a sufficient number of feature points using the algorithm proposed in~\cite{shi1994good}. 
Then it composes a feature points set and updates the coordinates of these feature points in following frames via optical flow. 

As shown in Figure \ref{fig:multitrack}, visual tracker generates an array called ``bitmap" to support multi-object tracking.
The length of bitmap equals to the number of feature points. 
Bitmap records the existence and group belonging of each feature point indicated by assigned values.
For each array element, ``0" means the corresponding feature point is lost during the tracking process, while positive value (e.g., object ID) indicates the point drops inside the contour of a specific object. After feature point extraction in the first frame, bitmap initializes all values to 1 to indicate the existences of these feature points, while the group belongings remain uncertain. 

Visual tracker extracts and tracks a group of feature points.  
It is a rare case for visual tracker to use the same set of feature points from the beginning to the end. Feature points regeneration may happen in two scenarios, namely, new objects are recognized by the server or  the number of feature points decreases to a certain threshold.
The former one indicates either that new objects have moved into the camera view, or the camera view has changed dramatically.  
The latter one is caused by some feature points lost during tracking, or by camera view changing as well.

The visual tracker extracts a new set of feature points from a camera frame when any of above mentioned conditions is met. This design ensures that any pair of consecutive frames always have enough number of corresponding feature points from a same set.
Therefore it enables uninterrupted object pose estimation.

\textbf{The object pose initialization and update pipeline}, working on top of the former pipeline, starts from the first offloading result. 
Taking any pair of frames as an example, if we knew the old pose of an object and the old locations of feature points within its contour, we can calculate the new corresponding metrics using geometric transformation. 

As shown in Figure \ref{fig:frames}, 30 camera frames compose one logical cycle of this pipeline. Camera frames can be divided into three categories based on their functionalities, i.e., \textit{key frames}, \textit{continuous frames} and \textit{result frames}. The visual tracker sends out a recognition request when receiving the first frame of each cycle, i.e., a key frame. Meanwhile, the tracker keeps a copy of the current feature points. It receives the recognition result at a later frame, i.e., a result frame. 
\sysname uses the preserved feature points from key frames and result frames to calculate the new poses of the recognized objects in result frames.

The visual tracker processes a target object with three steps: a) position initialization as described above; b) position tracking via geometric transformation between consecutive continuous frames; c) position update with new recognition results from latter result frames. 
We design this logical cycle for two reasons. 
First, new target objects can appear continuously within view. apparently, increasing the offloading frequency can shorten the cold start latency for recognizing new object, but the logical cycle time should be longer than offloading latency. We choose 30 frames ($\sim$ 1 second) as the cycle length considering the offloading latency under different networks.
Second, drift problem should always be considered during the tracking process. Otherwise the augmented contents will move away from right position gradually. With regular recognition requests, \sysname correct the positions of objects with recent recognition result to achieve higher accuracy.

The two pipelines cooperate and hide the offloading delay between the key frame and the result frame from users' perception. Our tracking method is not only especially suitable for MAR applications under edge offloading scenario, but also provides strong latency compensation for image recognition offloading in further-located cloud servers.

\subsection{Content Renderer}
\label{sec:contentRenderer}
Content renderer is the output module of \sysname, which uses a 3D graphics engine to render the virtual contents with 6DoF poses of objects. This ensures that virtual contents attach precisely to physical objects in a 3D world space. For each camera frame, the graphics engine updates the virtual contents based on the output of the visual tracker. For each object, the engine first calculates a homography to find the 2D transition between the reference image and the target image, then casts the 2D position into a 3D spatial pose.

\subsection{Network Module}

The network module is the messenger between the client and the server. UDP, as a connection-less protocol, saving the overhead of handshaking dialogues, becomes our choice. Packet losses may happen as UDP is an unreliable communication method, but our system tolerates packet loss naturally: edge recognitions happen periodically, and new recognition result can be received very soon in next cycle. Even if a reliable network protocol is used for visual task requests, the results could come late due to the re-transmission of lost packets, which is not acceptable in this real time system. 
\sysname uses non-blocking UDP in independent threads so that the client and server will not be blocked for network issues.

\section{Server design}
\label{sec:serverDesign}
In mobile AR systems, continuous visual processing in real time is nearly impossible on current devices. In \sysname framework, we use edge and cloud servers for those heavy computer vision tasks. Therefore the mobile and wearable clients can provide fluent AR experience and abundant contents. 

\subsection{Task Manager}
\label{sec:taskManager}
Upon launching an AR application, the mobile client will communicate with the server to initialize the edge service. 
On server side, task manager will create three threads for receiving requests, visual processing and sending results, respectively.

\subsection{Visual Task Processor}
\label{sec:visualtaskprocessor}

\begin{figure}[t!]
\center 
\begin{subfigure}[t]{0.23\textwidth}
    \includegraphics[width=\textwidth]{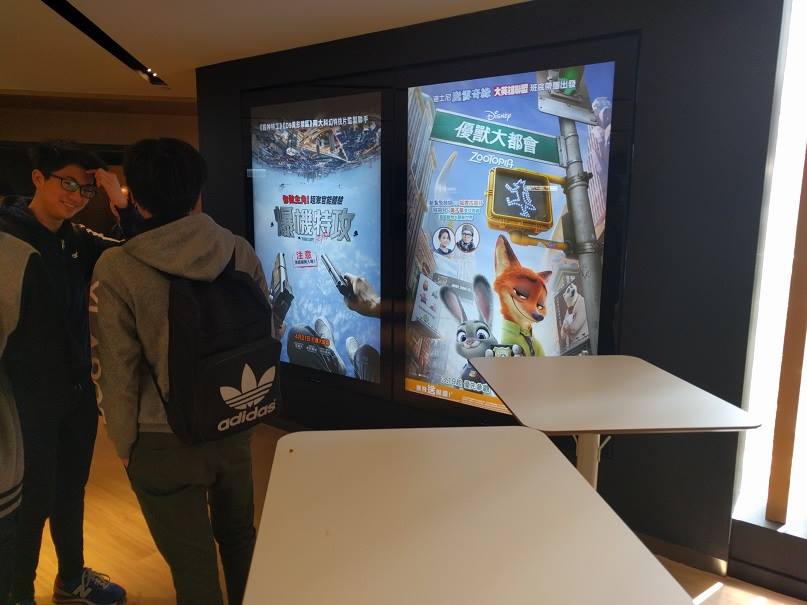}
    \caption{Camera frame input sent from the client to the server.}
    \label{fig:segmentation1}
    \end{subfigure}\hfill
\begin{subfigure}[t]{0.23\textwidth}
    \includegraphics[width=\textwidth]{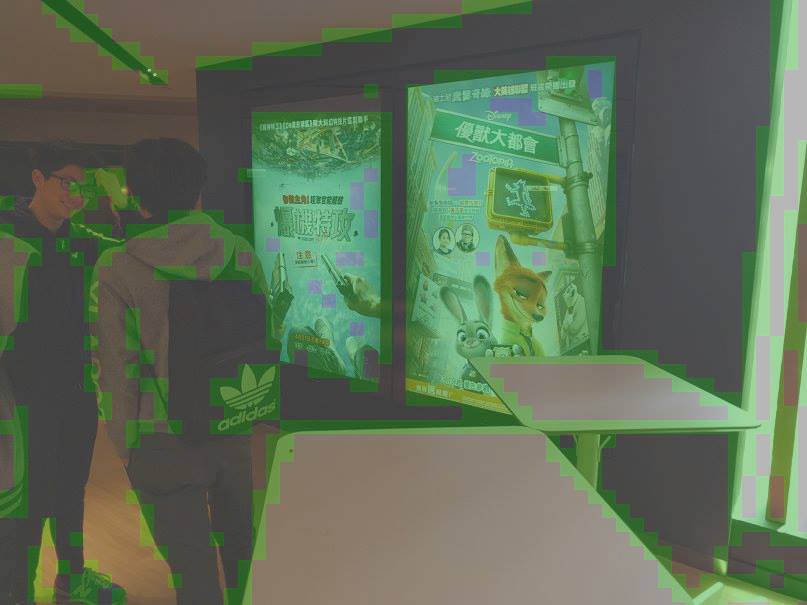}
    \caption{Areas with high pixel variance are highlighted.}
    \label{fig:segmentation2}
    \end{subfigure}
\begin{subfigure}[t]{0.23\textwidth}
    \includegraphics[width=\textwidth]{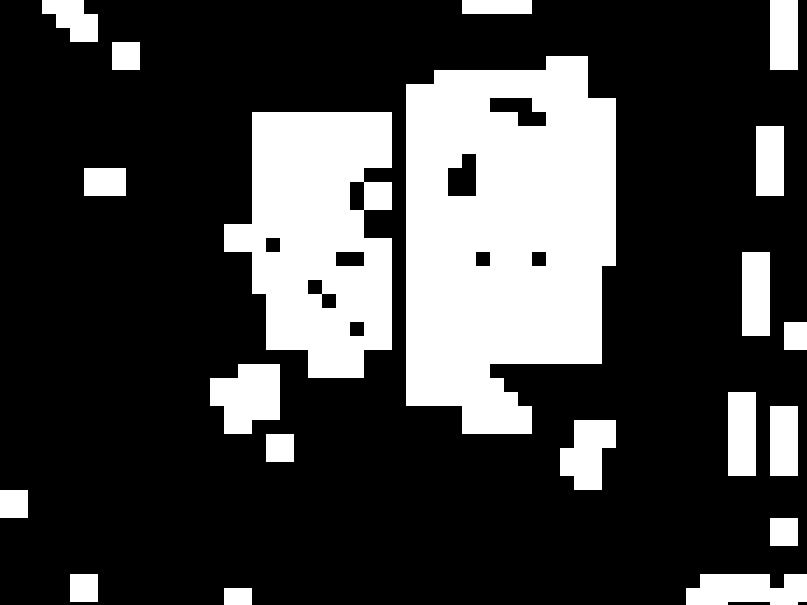}
    \caption{Identified areas before noise reduction as binary image.}
    \label{fig:segmentation3}
    \end{subfigure}\hfill
\begin{subfigure}[t]{0.23\textwidth}
    \includegraphics[width=\textwidth]{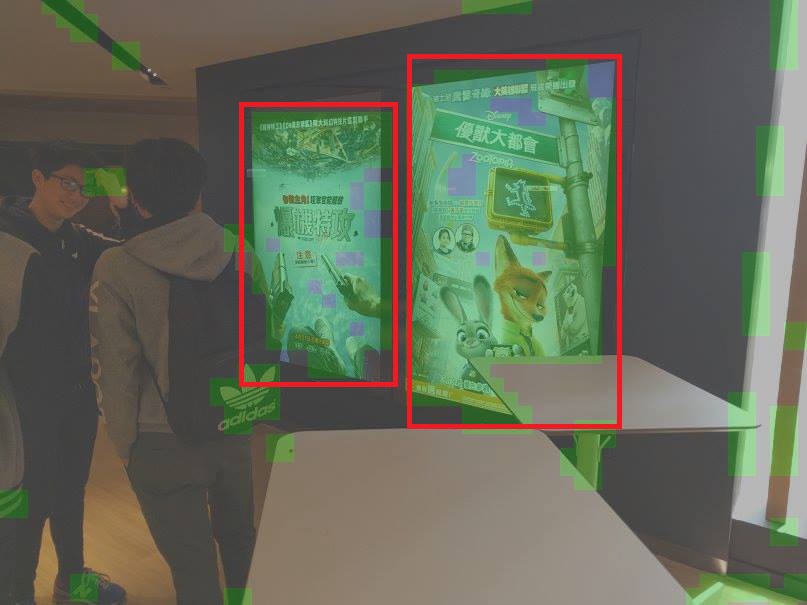}
    \caption{The objects of interest are identified and segmented.}
    \label{fig:segmentation4}
    \end{subfigure}
  \caption{Procedures of the image segmentation.}
  \label{fig:image_segmentation}
\end{figure}

Visual task processor fulfills visual requests by extracting desired information from camera frames through following steps.

\subsubsection{Image Segmentation}

Upon receiving an incoming camera frame from the client, the server starts to perform the image processing tasks. 
The processor segments the camera frame before retrieving information from it to reduce system load and speed up the process. The image segmentation identifies and extracts the segments in images that contain the image objects while removing the background. This effectively ensures that each query patch contains at most one reference image. 

The segmentation process is shown in Figure \ref{fig:image_segmentation}. 
We first apply Gaussian blur to the raw image to reduce noise. Then we apply a sliding window that iterates through the image, and compute the variance in pixel color values for each window. We flag the areas with variance higher than a constant threshold as positive ones, while the areas with low variance as negative ones since they are likely to be part of the background. We further clean up the resulting binary figure by applying an erosion followed by a dilation. The processed image contains polygons outlining the located objects. The system can then extract the objects by cutting the polygons from the raw image.

\subsubsection{Image Retrieval}

\begin{figure*}[t]
    \centering
    \includegraphics[width=0.7\textwidth]{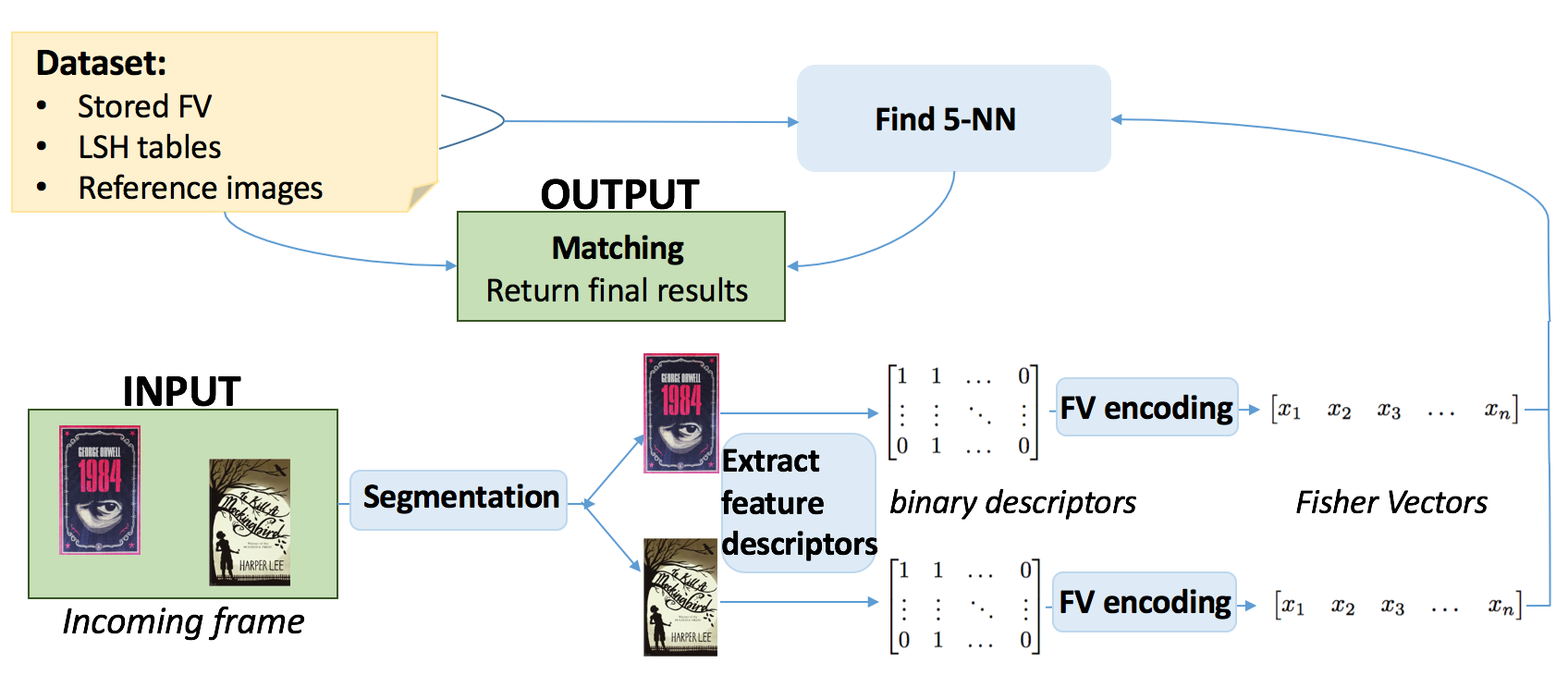}
    \caption{Image recognition pipeline. One frame is first segmented into multiple images, on which image retrieval techniques are applied to find nearest neighbors. Finally, template matching with neighbors verifies the results and calculates positions of the segmented images. }
    \label{fig:recognition}
\end{figure*}

Our system uses image features to retrieve corresponding image from the dataset.
The overall image retrieval pipeline is shown in Figure \ref{fig:recognition}. We first need to detect key points and extract corresponding features from each reference image to build the dataset. Then we need to concern the processing speed since the reference images (for building the dataset) and the query images (getting from camera frames) need to be handled through the same procedure. We use Multi-scale AGAST\cite{mair2010adaptive} detector and FREAK\cite{alahi2012freak} binary feature descriptors. We also adopt Fisher encoding built upon binary features for accurate and real-time performance. To do Fisher encoding on binary features, we first need to build a Bernoulli mixture model (BMM) upon the collection of all descriptors of the dataset images. After getting parameters of the BMM model, descriptors of an image can be encoded into a single Fisher Vector (FV). Encoded FVs will be L2 and power normalized, and all FVs of the dataset images will be stored and processed with Locality Sensitive Hashing (LSH) to create hash tables for faster retrieval.

After building up the dataset, the server can proceed the recognition tasks. 
It re-sizes the segmented individual patches simultaneously into scales similar with the reference images, and encodes them into FVs through the procedures mentioned above. The parameters for the BMM model used for query patch Fisher encoding is the same with the one used for encoding reference images. The LSH tables and stored FVs let the server find five nearest neighbors of each segmented patch, and the server select the correct one out of five using feature matching.

As the last step, feature matching verifies the result of image retrieval and calculates a 6DOF pose of the target. Feature matcher can find the corresponding feature pairs of each segmented patch using the feature descriptors from the patch and the nearest neighbors. We pick a neighbor as the recognition result and start pose estimation only if it has enough matches. We calculate the homography of the target within camera view and return simple corner positions to the task manager. The patch will be discarded without returning any result if there are not enough good matches for all five nearest neighbors in the dataset.

\section{Implementation and Evaluation}
\label{sec:evaluation}
The mobile client is implemented on the \textit{Android} platform. The mobile visual tracker is implemented with \textit{OpenCV4Android}\footnote{\url{http://opencv.org/platforms/android.html}} library, and the 3D renderer is a modified \textit{Rajawali}\footnote{\url{https://github.com/Rajawali/Rajawali}} library. 
We implement the edge server in C++ based on Linux platform. \textit{OpenCV}\footnote{\url{http://opencv.org}} library is used for multi-scale AGAST feature points detection and FREAK descriptor extraction. For building LSH tables and finding approximate nearest neighbor of FVs, we select \textit{FALCONN}\footnote{\url{https://falconn-lib.org}} library. The parallel processing relies on \textit{OpenMP}\footnote{\url{http://openmp.org}}.

\begin{figure}[t]
    \centering
    \includegraphics[width=\linewidth]{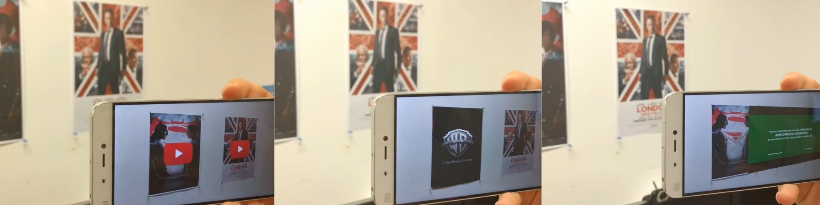}
    \caption{The PosterAR app running our \sysname framework. The app utilizes our framework to retrieve poster information and overlay movie trailers on the camera stream.}
    \label{fig:run}
\end{figure}

We develop PosterAR as a representative AR application to showcase our \sysname framework. Imagine the scenario when a user steps into a cinema, he/she does not know which movie to watch. Watching trailer can help the consumer select a movie. All he/she needs to do is to launch the PosterAR app, point the camera to the posters, then the trailer will automatically start to play. To further improve the QoE, the play buttons and movie trailers are strictly overlaid according to the homography of the posters, so that they are precisely coupled as if there is a virtual video wall.

Figure~\ref{fig:run} shows the running examples of PosterAR app. In the left part, the user is holding the phone against two movie posters and the app augments trailers on top of the posters. In the middle and right parts, movie trailers start to play in 3D after the user presses the play button. In this section, we collect relevant data from PosterAR to evaluate the performance of our framework.

\begin{figure*}[t]
    \centering
 \includegraphics[width=1\linewidth]{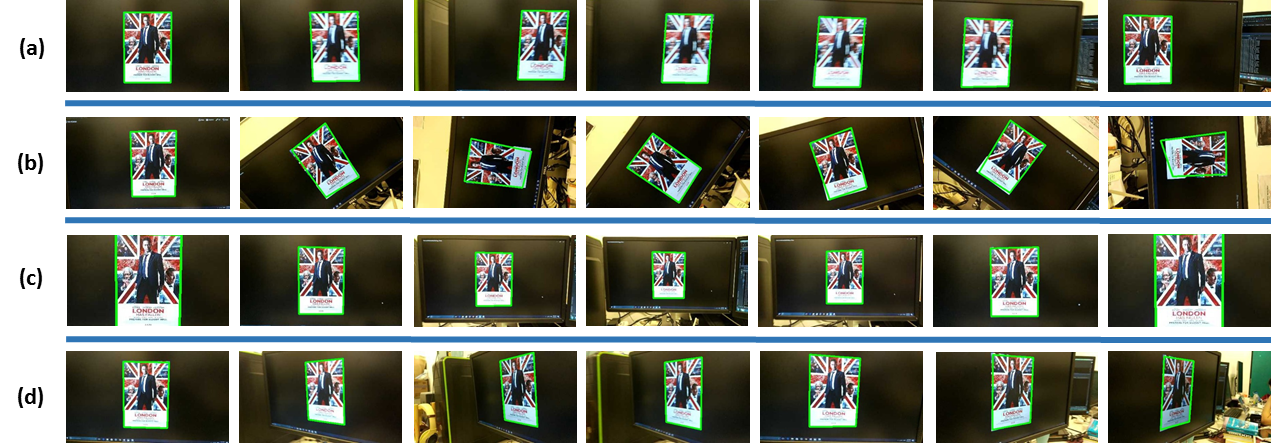}
    \caption{Four common scenarios of camera movement: (a) fast movement with motion blur; (b) camera rotate; (c) scaling caused by forward and backward movement; (d) tilt caused by changing view. The green edges are labeled with tracking results from the mobile client.}
    \label{fig:films}
\end{figure*}

\subsection{Experiment Setup}
\label{sec:experimentSetup}
On the client side, the PosterAR app runs on a Xiaomi MI5 with a quad core CPU and 4GB RAM. 
On the server side, we deploy servers on both a local PC and the Google Cloud Platform to respectively address the requirements of edge and cloud computing. The local PC is configured with an Intel i7-5820k CPU (6 cores @3.3GHz) and 32GB RAM. We create a WiFi Access Point on the PC to let it connect to the phone directly. This setup refers to the scenario in which the edge computing components collocate with the nearby LTE tower or local router within one hop distance from the mobile device.  The virtual machine on Google Cloud has 8 vCPUs and 32GB RAM.

We conduct our experiments with respect to mobile tracking performance, offloading latency, and image retrieval accuracy. 
In the end, we compare several performance metrics of edge and cloud recognition functionalities between \sysname and a leading commercial AR framework.

\subsection{Tracking Performance}
\label{sec:trackingPerformance}
Visual tracker is the key component of the \sysname framework, whose performance affects directly the user's perceived experience. In this section, we evaluate the performance of visual tracker with respect to the tracking accuracy and tracking speed.

\subsubsection{Tracking Accuracy}
An important indicator of the tracking performance is the quality of 6DoF tracking, which is crucial in providing a seamless AR experience. We design four common tracking scenarios to evaluate the performance: \textit{fast movement}, \textit{rotation}, \textit{scaling}, and \textit{tilt}. The results are shown in Figure \ref{fig:films}.

\begin{figure}[t]
\center 
\begin{subfigure}[b]{0.48\textwidth}
\center
    \includegraphics[width=.75\textwidth]{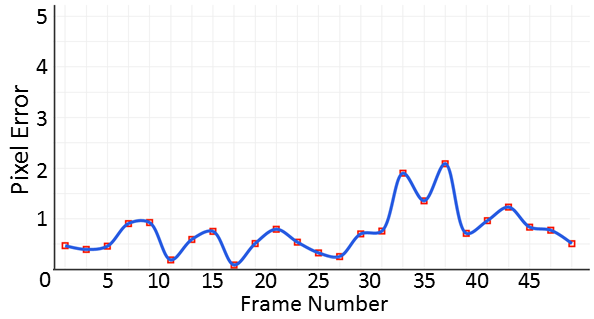}
    \caption{Fast Movement.}
    \label{fig:moveres}
    \end{subfigure}\hfill
\begin{subfigure}[b]{0.48\textwidth}
\center
    \includegraphics[width=.75\textwidth]{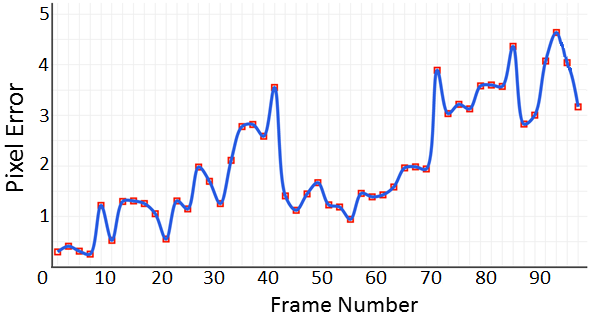}
    \caption{Rotation. }
    \label{fig:rotateres}
    \end{subfigure}
\begin{subfigure}[b]{0.48\textwidth}
\center
    \includegraphics[width=.75\textwidth]{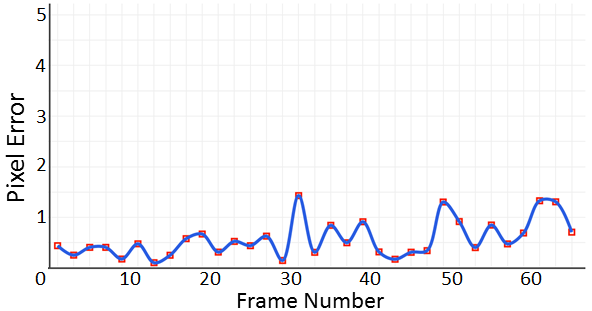}
    \caption{Scaling.}
    \label{fig:scaleres}
    \end{subfigure}    \hfill
\begin{subfigure}[b]{0.48\textwidth}
\center
    \includegraphics[width=.75\textwidth]{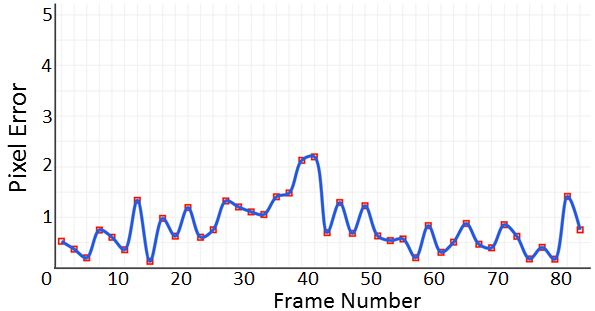}
    \caption{Tilt.}
    \label{fig:tiltres}
    \end{subfigure}    
   \caption{Tracking accuracy in pixel errors. Our visual tracker performs good in all four scenarios of camera movement with a pixel error below 2 pixels for fast movement, scaling and tilt, and below 5 pixels for rotation. }
  \label{fig:trackres}
\end{figure}

We record the positions of the poster's four corners in each frame given by the tracker, and obtain the ground truth by manually labeling corners for all frames. There are two common ways to measure the tracking accuracy: a) \textit{intersection over union (IOU)}, which considers the bounding boxes of both the tracking result and the ground truth, and it is defined as the proportion of intersection area within the union area of the two boxes; b) \textit{pixel error}, which is defined as the pixel distance on screen between the center of the tracking result and that of the ground truth. As our tracking method provides fine-grained 6DoF results, the bounding boxes are irregular quadrilaterals, whose intersection area and union area are complex to calculate. Instead, the center of each bounding box is more convenient to find, which is the intersection point of the two diagonals of the box. For this reason, we use \textit{pixel error} as the evaluation metrics in this part. The results are shown in Figure \ref{fig:trackres}, where the pixel errors in all scenarios are as low as 1 or 2 for most of the time. The worst case is rotation, but the pixel errors are still under 5. \added{Such low pixel error is within the precision expected by users given the high resolution of displays and their own tolerance to alignment issues. Note that in the case of see-through displays, the tolerance of users may be much lower and need to be validated through a user experiment.}

Also the rotation scenario shows a trend of drift, which is common in incremental tracking. However, our system handles the drift problem as the mobile client sends recognition requests periodically, and revises the position in tracking with the recognition results accordingly. This design guarantees that drift in tracking will last no longer than 30 frames in the worst case. 

\subsubsection{Tracking Speed}

The time consumption of tracking each frame is composed of four parts: time to down sample the frame, time to calculate the optical flow of feature points, time to estimate the poses of objects, and idle time waiting for next data. Since we use feature points based method in tracking, the amount of points would influence the performance and delay of tracker. To learn the impact of this factor, we measure the time consumptions of tracking camera frames with four different numbers of feature points, which are 60, 120, 180 and 240, correspondingly. For each case, we record the timestamps of different parts for 500 frames. 

\begin{figure}[t]
    \centering
    \includegraphics[width=0.4\textwidth]{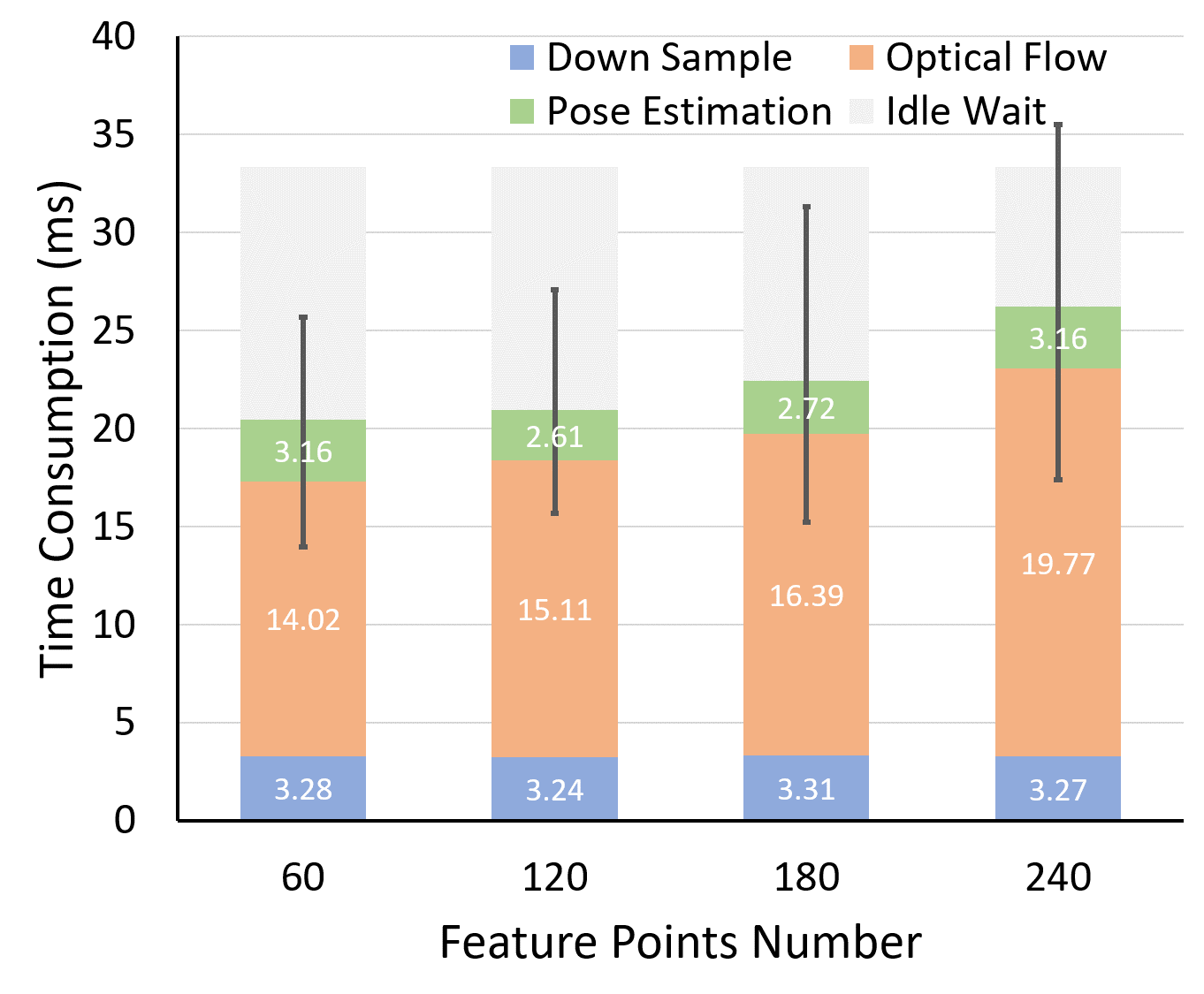}
    \caption{Time consumption for tracking one frame with different amount of feature points. The results are averaged over 200 runs with standard derivation shown in black line.}
    \label{fig:tracktime}
\end{figure}

The results are shown in Figure~\ref{fig:tracktime}. There is a clear trend that the time consumption of tracking one frame increases with the number of feature points, and the optical flow calculation part results in this time variance. To guarantee a 30 FPS performance of the tracker, the upper limit of the time consumption on each frame is 33.34ms. We can see that all four cases satisfy this requirement on average, and non-qualified only in worst cases with 240 feature points. We conclude that our visual tracker ensures a real-time performance with the amount of feature points less than or equal to 180.

\subsection{Offloading Latency}
\label{sec:latency}

Offloading latency is defined as the time period from the moment obtaining request camera frame data to the moment displaying offloading result, which is composed of three parts: delay of client, including the time to compose request and calculate new poses after receiving the result; delay of server, including the time to parse request and recognize image; delay of network, including the round trip time of the datagram packets.

In this section, we measure both the edge offloading latency and the cloud offloading latency with the PosterAR app. For each offloading scenario, 200 offloading tasks are generated and recorded.

\subsubsection{Latency Distribution}

\begin{figure}[t]
\center 
\begin{subfigure}[t]{0.48\textwidth}
    \includegraphics[width=0.8\textwidth]{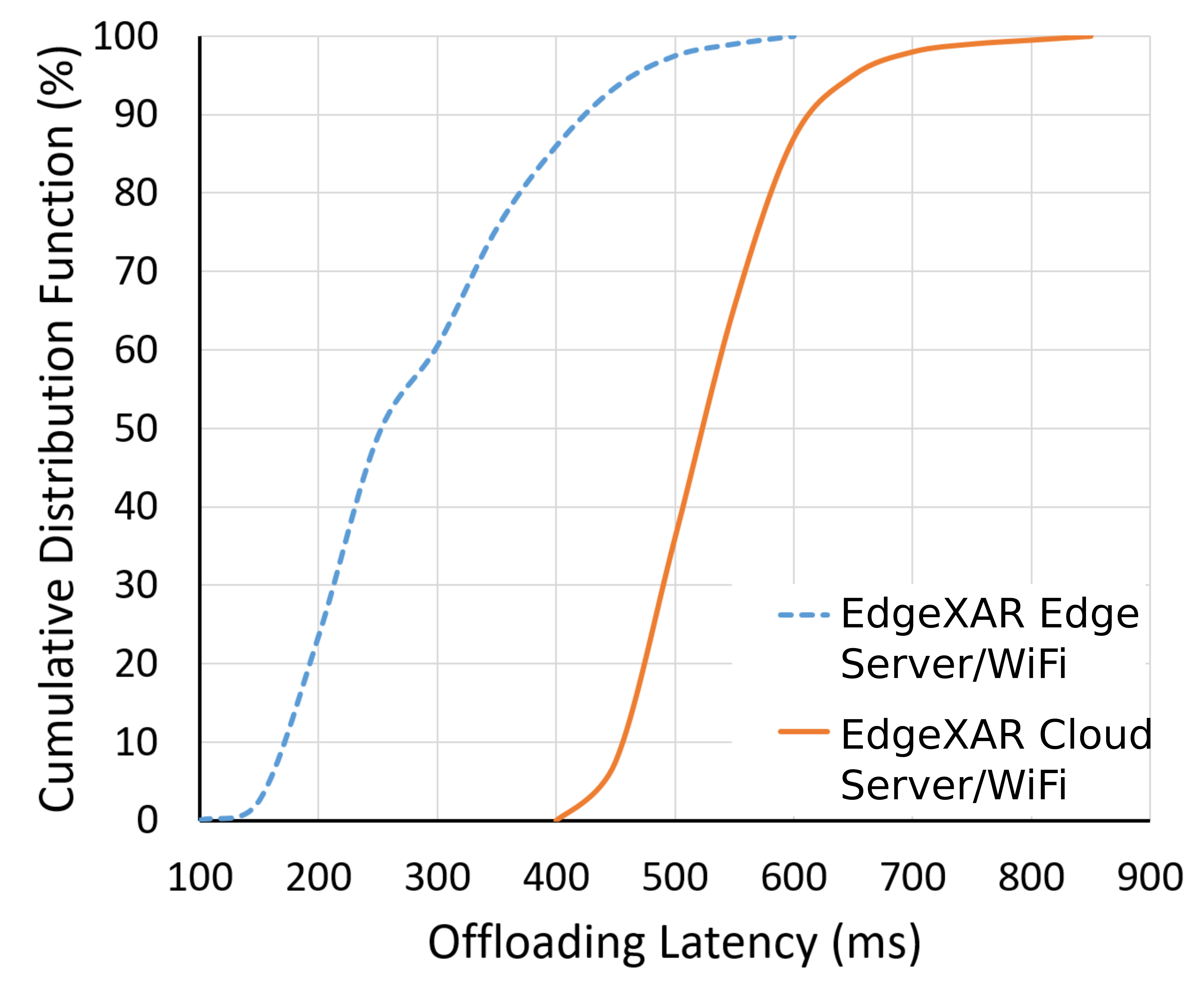}
    \caption{CDF of offloading latency for both edge offloading scenario and cloud offloading scenario.}
    \label{fig:offloadcdf}
    \end{subfigure}\hfill
\begin{subfigure}[t]{0.48\textwidth}
    \includegraphics[width=0.8\textwidth]{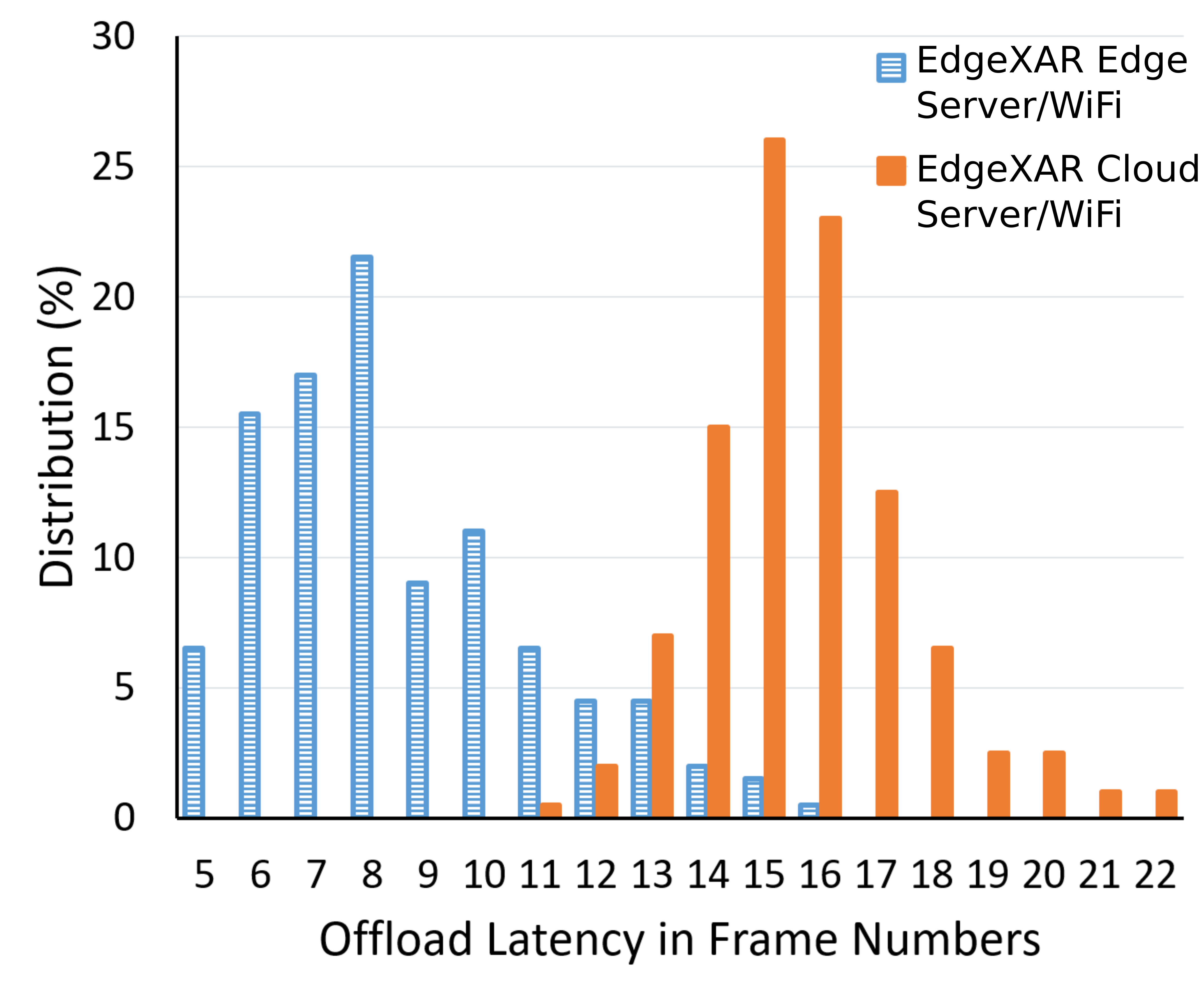}
    \caption{Distribution of offloading latency in passed frames. }
    \label{fig:offloadframe}
    \end{subfigure}
  \caption{Delay of offloading for both edge scenario and cloud scenario with 200 runs. Compared with cloud offloading scenario, edge offloading scenario displays a lower latency, as shown in both latency time and passed frames.}
  \label{fig:delay}
\end{figure}

Figure~\ref{fig:offloadcdf} is the CDF of overall offloading latency for two offloading scenarios, and edge offloading scenario display a lower latency than cloud offloading scenario. 60\% of the edge offloading tasks are finished before 300ms, and 97\% of them are finished before 500ms. On the contrary, only 35\% of the cloud offloading tasks are finished before 500ms, but all of them are finished before 850ms. (We also measure the latency of cloud offloading scenario under 4G connection in Section \ref{sec:cloudlatency}.) Figure~\ref{fig:offloadframe} is the distribution of offloading latency in passed frame numbers, which represents the number of frames that has passed (processed by visual tracker) before the offloading results are showed to user. For the edge offloading scenario, 60\% of the results are showed within 8 frames after the starting points of offloading, and 100\% of them are showed within 16 frames. Meanwhile, cloud offloading latency follows a normal distribution, and we come to the same conclusion as before that cloud offloading scenario has a longer latency than edge offloading scenario, as the mean of this ``normal distribution'' is 15 frames.

\begin{figure}[!tb]
    \centering
    \includegraphics[width=\linewidth]{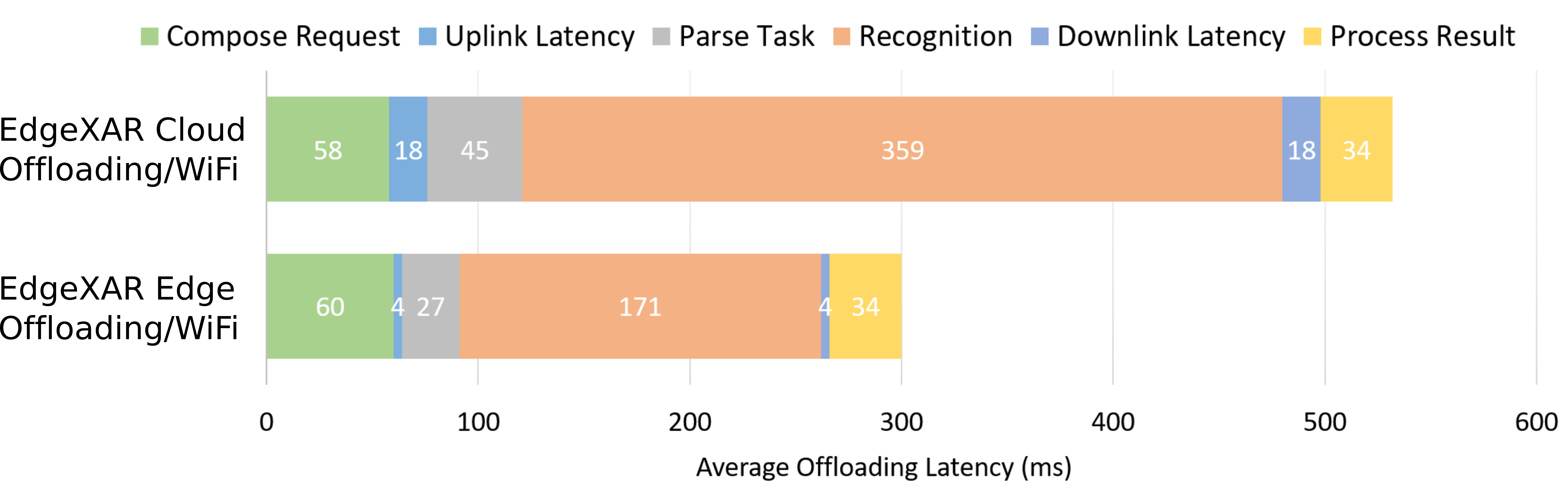}
    \caption{Components of latency for both edge offloading scenario and cloud offloading scenario. The results are averaged over 200 runs.}
    \label{fig:offloadlatency}
\end{figure}

Figure \ref{fig:offloadlatency} shows the components of latency for both scenarios. The part \textit{compose request} takes around 60ms, where encoding raw frame data into a small image file takes most of this time. The \textit{uplink latency} and \textit{downlink latency} parts in the edge offloading scenario are much shorter than that in cloud offloading scenario. \textit{Parse task} is the first part of the time spent on the server to parse the request as well as control the processing. \textit{Recognition} is the second part of the time spent on the server, including the time to segment the camera frame, find the nearest neighbor, match and verify the result. \added{Although we selected the VM to be as close as possible to the machine we use at the edge, we observe a significant discrepancy in the processing time on the server between the two use cases. For both \textit{parse task} and \textit{recognition}, the edge server processes faster than the cloud server, showing that the edge server has a better processing capability. Such discrepancy is most likely due to the different processing architectures between a VM in the cloud and a discrete CPU running the operations on the edge server.} \textit{Process result} is the time that the client processes the result and displays it to the user.

From the requirements of visual tracker, the whole offloading procedure should finish within 30 frames (1 second). The results from figure \ref{fig:offloadlatency} prove that our offloading pipeline fulfills this requirement under both edge and cloud offloading scenarios. 

\subsection{Image Retrieval Accuracy}
\label{sec:imageRetreivalAccuracy}

\begin{figure}[t]
    \centering
 \includegraphics[width=0.35\textwidth]{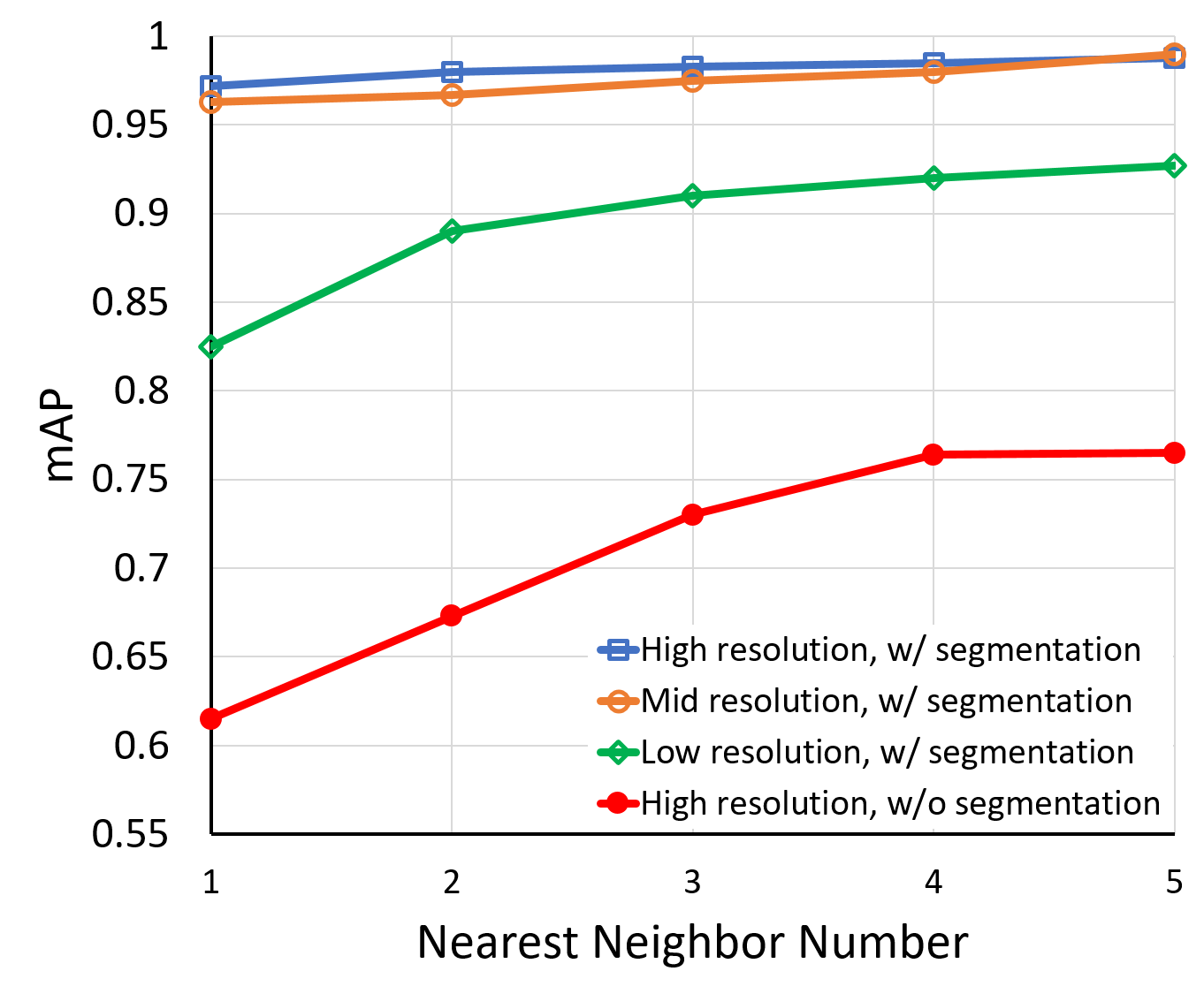}
    \caption{Image retrieval mean Average Precision (mAP) for matching 1-5 nearest neighbors (NN) with or without segmentation under different scales.}
    \label{fig:map}
\end{figure}

We collect several hundreds of movie posters online, combining with some data from Stanford Mobile Visual Search Dataset~\cite{chandrasekhar2011stanford} to form 1000 reference images. Query sets are photoed by Mi5's back camera, together with several query sets from~\cite{chandrasekhar2011stanford}.

If the query images are passed directly into the feature extraction stage without segmentation, then the resulting retrieval mean Average Precision (mAP) is around 0.76 for top-5 NN (nearest neighbor), and around 0.61 if we only check one most nearest neighbor, as shown by the ``without segmentation'' line in Figure~\ref{fig:map}. The relative low mAP is mainly caused by multi-target existence within one query image, or random scattered noise in the background.

Given good segmentation results, the retrieval can achieve good accuracy for top-5 NN. In our experiment, all patches will be re-sized into a 400$\times$400 resolution. From Figure \ref{fig:map} we can see that in most cases, we only need to examine the most nearest neighbor during the matching step to get the correct result, and the expectation of examined neighbor number is close to 1. The high-resolution line corresponds to a patch size of at least 400$\times$400 pixels. The mid-resolution line corresponds to having segmented patches with size around 200$\times$200 pixels, and the small-resolution line corresponds to patch size of 100$\times$100 pixels. During our experiment, more than 90\% segmented incoming frames can correctly retrieve their corresponding images in the dataset.

\subsection{Performance Comparison with Vuforia}
\label{sec:overallPerformance}

\textit{Vuforia} is a leading commercial AR framework that supports cloud recognition, and \sysname focuses on cloud-based AR solutions. Therefore, we compare several performance metrics of cloud recognition functionalities between \sysname and Vuforia. 

To setup the experiment, we download the official CloudReco app from Vuforia. This app performs image recognition tasks by offloading, receives results and displays augmented contents to user. 
On the Android side of Vuforia, we do essential instrumentation on CloudReco to log analysis data and influence the performance as little as possible. On the server side of Vuforia, we create a image database with their web services. Since Vuforia supports only one-by-one image uploading, we create a database with only 100 images, which is a portion of the dataset used for evaluating \sysname. For our PosterAR app, we use the previous setup mentioned before, and the cloud server hosted on Google Cloud Platform is used to pursue a fair comparison. Both WiFi and 4G connections of the phone are used to evaluate the performance of the two frameworks under different network conditions.

\subsubsection{Data Usage}
\label{sec:datausage}

Data transmission is always involved in offloading, so we want to compare the data traffic usage for the two frameworks. Through our network monitoring of the CloudReco app, we find out that cloud recognitions in Vuforia framework are implemented with Https requests. With this knowledge, we setup a man-in-the-middle (MITM) proxy using Fiddler8 and decrypt all the recognition requests and responses of the CloudReco app. Our key finding is that each request body contains a JPEG file of the
current camera frame, and each response body contains another JPEG file of the recognized original image from database. Since there are only 100 images in our created Vuforia demo database, we use this 100 images to trigger cloud recognition tasks on both frameworks. The average request sizes of the PosterAR app and the CloudReco app are 11.58KB and 16.24KB correspondingly, as both of them are sending compressed camera frame to the cloud. However, the average result sizes of the two apps differ hugely. In \sysname, the results contain only coordinates of the bounding
boxes and other identity information of the images, so we make
the result a fixed size of only 400 bytes. In Vuforia, the average size
of offloading results is 33.48KB, showing that Vuforia consumes
much more data traffic than \sysname in each offloading task.

\begin{figure}[t]
\center 
\begin{subfigure}[t]{0.48\textwidth}
    \includegraphics[width=0.8\textwidth]{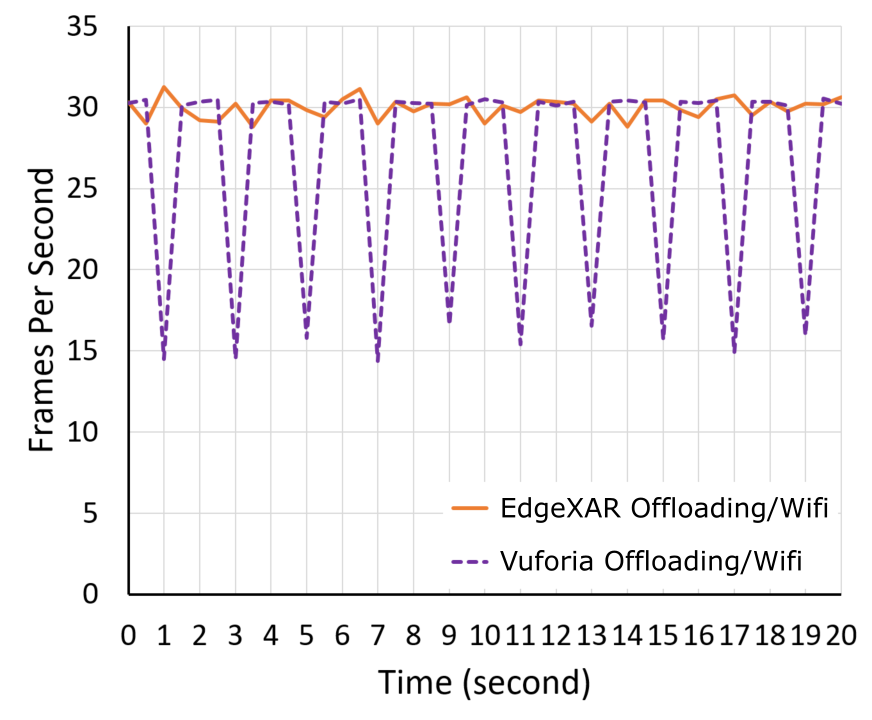}
    \caption{FPS of apps with offloading built with \sysname and Vuforia.}
    \label{fig:fps}
    \end{subfigure}\hfill
\begin{subfigure}[t]{0.48\textwidth}
    \includegraphics[width=0.8\textwidth]{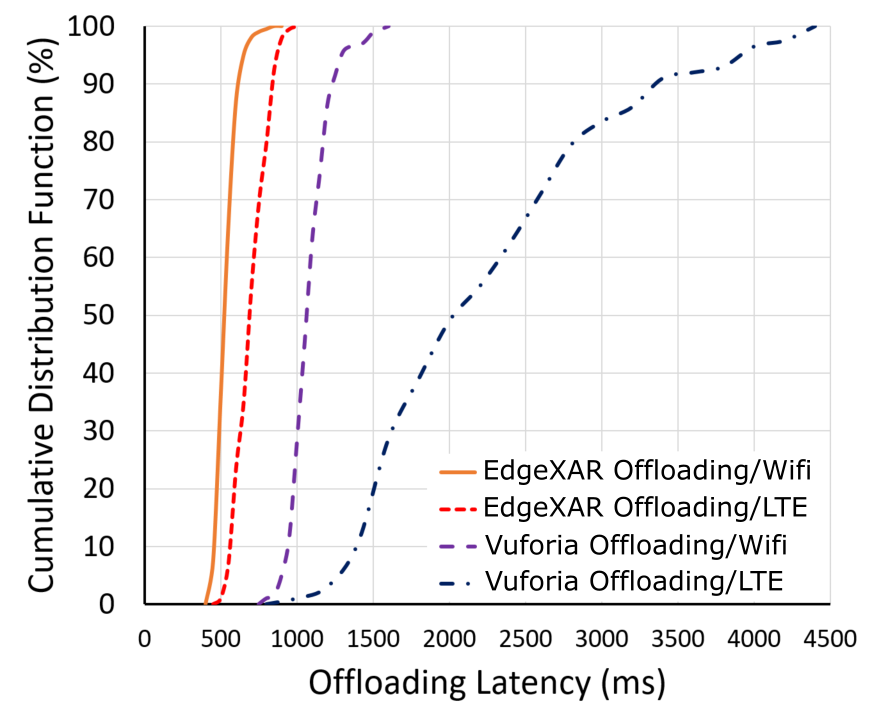}
    \caption{CDF of offloading latency for both \sysname and Vuforia with 200 runs.}
    \label{fig:vuforialatency}
    \end{subfigure}
  \caption{Performance comparison of \sysname and Vuforia in terms of FPS and offloading latency.}
  \label{fig:performance}
\end{figure}

\subsubsection{Runtime FPS}
\label{sec:fps}
FPS reflects the runtime performance of the apps, which has straightforward influence on the user perceived experience. Offloading procedure is an extra burden onto the  mobile device besides frame-by-frame tracking, so we want to learn how would offloading influence the FPS. 

The offloading procedure of the PosterAR app is periodical, but the CloudReco app would trigger an offloading task when there is no recognized object in the current frame. Imagine a scenario where the user scans through a series of posters on the wall, and periodical offloading happens in both frameworks. To simulate this scenario, we make a slideshow of the images on the screen and point the phone's camera to the screen when the two apps are launched. The slideshow has a fixed change interval of 2 seconds, and we record the runtime FPS of both apps. A typical fragment of the result is shown in figure \ref{fig:fps}. We can see that the FPS of the CloudReco app will decrease significantly upon receiving a result, and we perceive obvious non-smooth frames during the experiment. In comparison, the PosterAR app runs at a much stable FPS with little jitters around 30. As the offloading result of Vuforia is the original image, feature extraction and matching are executed on the client, which are computationally intensive for mobile devices that decrease the FPS significantly.  

\subsubsection{Offloading Latency}
\label{sec:cloudlatency}
We use the same scenario in measuring FPS to trigger 200 offloading tasks for both \sysname and Vuforia under WiFi and 4G connections, and the offloading latency results are shown in figure \ref{fig:vuforialatency}. We have already seen the CDF of \sysname cloud offloading latency under WiFi connection in previous section, and the CDF of that latency under 4G connection has a similar pattern with it, with slightly longer latency but still guaranteed to finish within 1 second. These latency results of \sysname cloud offloading prove that our periodical offloading design works with cloud servers under both WiFi and 4G connections. 

On the contrary, Vuforia cloud offloading has a much longer latency, especially under 4G connection. For Vuforia cloud offloading under WiFi connection, only 30 percent of the offloading tasks are finished with 1 second, and the worst case takes around 1.6 seconds. It is worth to mention that there is not much difference in the network accessing latency of Vuforia and \sysname cloud servers, with RTTs of 48 ms and 36 ms, respectively under WiFi connection. Vuforia cloud offloading under 4G connection has a noticeable poorer performance, with only 50 percent of the offloading tasks are finished within 2 seconds, and the worst case takes as long as 4.4 seconds. This poor performance of Vuforia under 4G possibly comes from the combination of large result size and lossy 4G connection, which turns into a long cold-start latency. 	

\added{It is important to note that the fifth generation of mobile networks promises latency below 5\,ms for the enhanced Mobile Broadband (eMBB) service class~\cite{38913}. As such, we expect the offloading latency in 5G to be similar to our WiFi experimental condition.}

\subsubsection{Multi-target Recognition Support}
\label{sec:multitarget}
\sysname framework supports multi-target recognition from the design of mobile visual tracker and cloud image recognition pipeline, and a running example of recognizing multiple targets is showed in figure \ref{fig:run}. In \sysname framework, most of the computation burden of recognizing multiple target is offloaded to server, and the mobile client only processes simple tasks like calculating the bounding boxes with offloading results. On the contrary, the CloudReco app does not support multi-target recognition, and only one target within camera view is recognized every time. Even if multi-target cloud recognition is supported under the current Vuforia offloading design, multiple original images would make the results quite huge, and the heavy feature extraction and template matching processes for multiple images on mobile would ruin the user experience. As a result, \sysname will enable users to interact with multiple interactive objects in AR.


\added{\section{Discussion}

\subsection{Application of \sysname and Usecases}

In this paper, we demonstrate \sysname through a single movie poster recognition application. However, \sysname provides functions to develop more complex applications. \sysname benefits all applications requiring fine alignment of digital content over the physical world. Typical use cases may involve sensor information display in smart homes. In physical houses, it is common practice to regroup all control functions within the same location. For instance, all light switches are often grouped next to a room's entry or exit doors, often with the control panels for thermostats. Similarly, controls for house alarm, intercom, and outdoors domotics are often located next to the front door. In such a scenario, providing accurate alignment of the digital controls with their physical counterpart is critical to avoid user confusion when displaying information and preventing erroneous manipulation. For instance, a misalignment with a smart intercom control placed next to a smart home alarm control may lead to the user accidentally setting up the alarm while trying to answer the intercom. \sysname also has applications in pervasive AR, where users wear headsets in the street and overlay content over buildings' facades. The lightweight headsets used in mobile scenarios are often even less powerful than smartphones, requiring to leverage the power of remote machines. The object detection, perspective estimation and rendering capabilities of \sysname enable displaying content over the building facades, for instance, over storefronts. Meanwhile, the latency compensation technique allows a smooth user experience even on highly variable mobile networks, especially 5G.

\subsection{AR application Development with \sysname}

We implement \sysname as a Java library for the Android mobile operating system. Similar to other AR frameworks, this library exposes the typical functions for AR applications, including frame preprocessing, feature extraction, object detection, object recognition, template matching, object tracking, and annotation rendering as defined in Figure~\ref{fig:arpipeline}. The \sysname framework provides a level of abstraction hiding how these functions are executed to the developer. As such, the developer can call the functions in the application code without considering whether the function will be executed on-device or in the cloud. Functions may experience longer latency due to the network component (see Section~\ref{sec:cloudlatency}). As such, we implement them as asynchronous callbacks, leaving to the developer control over the execution order of the AR pipeline.

It is important to note that the implementation of \sysname was meant to abstract the remote execution of functions over the network. As such, the server-side of \sysname is not meant to be directly modified by typical AR application developers. \sysname's server thus operates as a standalone application. For more advanced developers, adding new features would require adding a new module to the server-side and implement the abstraction at the client-side.

\subsection{Limitations and Future Works}

In this paper, we introduced \sysname, a framework for remote execution of AR functions with latency compensation. \sysname addresses the added network transmission latency through a technical angle. The proposed latency compensation technique uses local object tracking to recompute the results of object recognition returned by the server. As such, we evaluate \sysname through technical measurements, and show that the average error is often below what users can perceive on handheld devices. However, for extreme use cases (high latency, unexpected losses, complex scenes with multiple objects), such compensation techniques may not be sufficient. In future works, we would therefore focus on evaluating the limits of latency compensation techniques on the user experience through user studies. Such studies would allow us to design compensation solutions adapted to the user's perception of the system, and focus on the primary pain points of the AR experience. From a technical perspective, we will also consider improving  the optical flow-based tracking which is not resistant to occlusion. We plan to integrate sensor fusion and other occlusion resistant tracking methods to handle this problem. And the performance of our current image segmentation method is unstable on complex backgrounds, thus we are considering of integrating multiple methods to overcome this problem.
}

\section{Conclusion}
\label{sec:conclusion}
In this paper, we presented \sysname, an edge-based AR framework which solves large-scale multiple image recognition and real-time augmenting problems on mobile and wearable devices. We proposed an 
highly flexible tracker running on mobile to hide the offloading latency from user's perception, and a multi-object image retrieval pipeline running on server. We further showed that our framework can support user-centric AR, with the prominent features of high tracking accuracy, high recognition accuracy, low overall offloading latency, and high runtime performance.
Our evaluation showed that the AR app built with \sysname performs effectively in terms of seamless AR experience and robustness. \sysname performs tracking in less than 25\,ms for up to 240 feature points, and presents a low (1\textasciitilde 2) pixel error, allowing to efficiently track multiple objects over a scene at 30\,FPS, and hide an offloading latency of up to 600\,ms.
 Compared to a major commercial framework, \sysname provides more stable performance, with a framerate between 29 and 31\,FPS, transmits 87\% less data, halves the offloading latency in Wifi, and reduces it by 70\% in LTE, while supporting multi-target recognition.

\section*{Acknowledgment}
  This work was supported in part by 5GEAR project (Decision No. 318927) and the FIT project (Decision No. 325570) funded by the Academy of Finland.

\bibliographystyle{ACM-Reference-Format}
\bibliography{main}

\end{document}